\documentclass[11pt,twoside]{article}

\newcommand{\sindex}[1]{{\it #1}}
\newcommand{\oindex}[1]{{\rm #1}}
 
\usepackage{asp2004} \usepackage{epsf} 
\usepackage{lscape}

\begin{document}
\title{A Uniform Set of Optical/NIR Photometric Zero Points to be Used with CHORIZOS}   
\author{J. Ma\'{\i}z Apell\'aniz}    
\affil{Space Telescope Science Institute, Baltimore, MD 21218, USA \\
Instituto de Astrof\'{\i}sica de Andaluc\'{\i}a, Granada 18008, Spain (current)}  
 
\begin{abstract}  
I have recently combined HST/STIS spectrophotometry with existing photometric data to analyze the calibration of 
three standard optical photometry systems: Tycho-2 $B_{\rm T}V_{\rm T}$, Str\"omgren $uvby$, and 
Johnson $UBV$. In this contribution I summarize those results, present new ones for 2MASS $JHK_s$, and combine them with 
recent literature results to generate a uniform set of zero points for six photometric systems, the above 
mentioned plus Cousins $RI$ and SDSS $ugriz$. With the exception of the latter system, the zero 
points use the new Vega spectrum presented at this meeting by Ralph Bohlin. I also discuss
the implementation of these results in CHORIZOS, a Bayesian photometric code that compares multi-filter
observational data with spectral energy distributions to solve the inverse problem of finding the models which
are compatible with the observations. 
\end{abstract}

\section{Introduction}

The inverse photometric problem in astronomy can be defined in the following way: one observes a 
series of magnitudes $m_{{\rm obs},p}$ in a series of filter passbands (denoted here by the $p$ index), each one of 
them defined by a total-system dimensionless sensitivity function $P_p(\lambda)$, and compares them with a 
family of \sindex{spectral energy distribution}s (SED, denoted here by the $s$ index) $f_{\lambda,s}(\lambda)$ to find 
out which of the SEDs is compatible with the observed magnitudes. The first required step in that process is to 
compute the synthetic magnitudes $m_{r,p}$ in the $r$ system for each SED in each passband $p$.
For a photon-counting detector, the formula is:

\begin{equation}
m_{r,p}[f_{\lambda,s}(\lambda)] = 
 -2.5\log_{10}\left(\frac{\int P_p(\lambda)f_{\lambda,s}(\lambda)\lambda\,d\lambda}
                         {\int P_p(\lambda)f_{\lambda,r}(\lambda)\lambda\,d\lambda}\right)
                       + {\rm ZP}_{r,p}.
\label{photon}
\end{equation}

	A magnitude system $r$ is defined by a reference SED $f_{\lambda,r}(\lambda)$ and a series of (relative) 
zero points ZP$_{r,p}$ for each filter. In principle, one can adjust the $f_{\lambda,r}(\lambda)$ in such a way 
that ZP$_{r,p}$ = 0.0 for all filters, which generates the default system for that SED. Examples are given in Table
1. Some magnitude systems (e.g. Str\"omgren+Vega), however, use non-zero ZP$_{r,p}$ by definition. 

	It is commonly found when one attempts to use a magnitude system with zero ZP$_{r,p}$ that,
after the data has been fully processed, there are small offsets between observed and synthetic values
(typically in the hundredths or thousands of magnitudes) that have to be corrected\footnote{Given the unintended
nature of this effect, some authors refer to these quantities as zero-point offsets instead of zero points.}.
There are two approaches to deal with this issue: the first one, used e.g. for \sindex{HST photometry}, is to leave the
synthetic magnitudes unchanged and modify $m_{{\rm obs},p}$ by changing the calibration parameters that are used
to convert from instrumental counts to observed magnitudes. This leaves the zero points (as defined here) as exactly zero, e.g. 
ZP$_{{\rm HST+Vega},p} = 0.0$ for all filters. Two problems with this approach are that the observed magnitudes of
a given constant object (e.g. a calibration star) change when the reference SED is modified and that it would be hard
to apply to systems with long histories (e.g. Johnson $UBV$). An alternative approach, which will be followed here,
is to calculate the non-zero values of ZP$_{r,p}$, leaving the published observed magnitudes fixed and 
introducing changes in the synthetic ones. This is more practical way to deal with historical systems but it means that,
for example, the synthetic magnitudes in the Johnson+Vega system\footnote{That is, the 
values that should be compared with the observed ones.}, $m_{{\rm Johnson+Vega},p}$, where
$p=U,B,V$, are not the same as those calculated in the default \sindex{Vega system} (also called 
VEGAMAG), $m_{{\rm Vega},p}$, due to the existence of zero points. This one is also a more practical approach for large-scale surveys
(Tycho-2, SDSS, 2MASS\ldots) where the magnitudes need to be published as soon as possible and it is not until a few years 
later that accurate zero points are calculated. As an example, synthetic SDSS magnitudes, $m_{{\rm SDSS+AB},p}$, where 
$p=u,g,r,i,z$, are reported in an AB-based system and require a non-zero 
ZP$_{{\rm SDSS+AB},p}$ in order to be accurately compared with observed magnitudes.

\bigskip

\centerline{{\bf Table 1.} Default magnitude systems (ZP$_{r,p} = 0.0$ for all $p$).}

\medskip

\centerline{
\begin{tabular}{lllll}
\hline
\hline
System     & \multicolumn{4}{c}{Reference spectral energy distribution}                                 \\
\hline
Default ST & $f_{\lambda,{\rm ST}}$   & = & $3.63079\cdot 10^{-9}$  & erg s$^{-1}$ cm$^{-2}$ \AA$^{-1}$ \\
Default AB & $f_{\nu,{\rm AB}}$       & = & $3.63079\cdot 10^{-20}$ & erg s$^{-1}$ cm$^{-2}$ Hz$^{-1}$  \\
VEGAMAG    & $f_{\lambda,{\rm Vega}}$ & = &\multicolumn{2}{l}{Vega spectrum}                            \\
\hline
\end{tabular}}

\bigskip

	The relationships between different magnitude systems are easily derived from Eqn~\ref{photon}.
We start with the magnitudes of \oindex{Vega} in the default ST and AB systems, which are given by:

\begin{equation}
m_{{\rm ST},p}({\rm Vega}) =
 -2.5\log_{10}\left(\frac{\int P_p(\lambda)f_{\lambda,{\rm Vega}}(\lambda)\lambda\,d\lambda}
                         {\int P_p(\lambda)f_{\lambda,{\rm ST}}           \lambda\,d\lambda}\right),
\label{VegaST}
\end{equation}

\begin{equation}
m_{{\rm AB},p}({\rm Vega}) =
 -2.5\log_{10}\left(\frac{\int P_p(\lambda)f_{\lambda,{\rm Vega}}(\lambda)\lambda\,d\lambda}
                         {\int P_p(\lambda)(cf_{\nu,{\rm AB}}            /\lambda) d\lambda}\right).
\label{VegaAB}
\end{equation}

	$m_{{\rm ST},p}({\rm Vega})$ and $m_{{\rm AB},p}({\rm Vega})$ can then be used to transform
from an X+Vega magnitude system (where X = \sindex{Johnson}, \sindex{Cousins}\ldots) into ST or AB magnitudes by applying:

\begin{equation}
m_{{\rm ST},p} = (m_{{\rm X+Vega},p} - {\rm ZP}_{{\rm X+Vega},p}) + m_{{\rm ST},p}({\rm Vega}),
\label{mST}
\end{equation}

\begin{equation}
m_{{\rm AB},p} = (m_{{\rm X+Vega},p} - {\rm ZP}_{{\rm X+Vega},p}) + m_{{\rm AB},p}({\rm Vega}).
\label{mAB}
\end{equation}

	Similarly, if we assume ${\rm ZP}_{{\rm X+Vega},p} = {\rm ZP}_{{\rm X+ST},p} = {\rm ZP}_{{\rm X+AB},p}$, we can transform from
an X+ST or X+AB magnitude system into an X+Vega system by using:

\begin{equation}
m_{{\rm X+Vega},p} = m_{{\rm X+ST},p} - m_{{\rm ST},p}({\rm Vega}),
\label{mVegaST}
\end{equation}

\begin{equation}
m_{{\rm X+Vega},p} = m_{{\rm X+AB},p} - m_{{\rm AB},p}({\rm Vega}).
\label{mVegaAB}
\end{equation}

	Some authors list the zero-magnitude $f_\lambda$ or $f_\nu$ for a given magnitude system and filter. Those quantities are 
easily derived from Eqns.~\ref{mST}~and~\ref{mAB}:

\begin{equation}
f_{\lambda,p}(0\; {\rm mag}) = f_{\lambda, {\rm ST}}\cdot 10^{-0.4[m_{{\rm ST},p}({\rm Vega})-{\rm ZP}_{{\rm X+Vega},p}]},
\label{fl0}
\end{equation}

\begin{equation}
f_{\nu,p}(0\; {\rm mag}) = f_{\nu, {\rm AB}}\cdot 10^{-0.4[m_{{\rm AB},p}({\rm Vega})-{\rm ZP}_{{\rm X+Vega},p}]}.
\label{fn0}
\end{equation}

	It has not been until recently that the quality and stability of the observational data and
calibrations in astronomy has been good enough to allow for a measurement of the photometric zero points of a 
magnitude system with an accuracy of $\le 1\%$. Such an accuracy is necessary to solve the inverse photometric
problem without introducing significant systematic errors. In this contribution I summarize the recent work that has attained that goal and
I generate a uniform list of NIR/optical photometric zero points.

\section{A new Vega spectrum}

\sindex{Vega} is the reference SED for many magnitude systems. However, its spectrum has been traditionally difficult to measure with high 
accuracy over large wavelength ranges. For example, two of the frequently used SEDs for Vega, those of \citet{Coheetal92} [used to link optical
and IR] and \citet{BohlGill04a} [optically-based and extended into the IR using a 9550 K \sindex{Kurucz model}] differ by 2--3\% in flux in the NIR. 
Part of the difficulty in obtaining an accurate spectrum is due to the fact that Vega is a near-pole-on fast rotator \citep{Peteetal06,Aufdetal06}, with a 
temperature range between 7900 K and 10\,150 K over its surface.

	\citet{Bohl07} presents in these proceedings a new FUV-to-IR HST/STIS-based Vega spectrum that uses a revised CTE correction
\citep{Goudetal06}, which is necessary due to the radiation damage that CCDs suffer in space. The new 
SED uses a 9400 K Kurucz model for long wavelengths that improves the overall agreement with the \citet{Coheetal92} model in the NIR to better
than 1\%. There are still some discrepancies between the two, some due to the different absolute calibration for Vega (\citealt{Mege95} vs.
\citealt{Haye85}) and the rest to the different spectral resolution and version (i.e. year) of the used Kurucz models. Nevertheless, the agreement is 
remarkable (and unprecedented) enough to prompt the usage of this new Vega SED\footnote{Available from
{\tt ftp://ftp.stsci.edu/cdbs/cdbs2/current\_calspec/alpha\_lyr\_stis\_003.fits} .} as a basis for the calculation of accurate zero points. For 
that reason, I have adopted it for CHORIZOS \citep{Maiz04c} starting with version 2.1.3 as well as for the zero points reported in this
contribution.

\section{Results from STIS spectrophotometry}

In two recent papers, \citet{Maiz05b} and \citet{Maiz06a}, I have combined HST/STIS spectrophotometry with existing space-based 
(for Tycho-2 $B_{\rm T}V_{\rm T}$) and ground-based (for Str\"omgren \sindex{$uvby$} and \sindex{Johnson $UBV$}) photometry to test the sensitivity curves 
of those three systems and derive the corresponding zero points. Those papers use two samples, the Next Generation Spectral Library (NGSL; 
\citealt{Gregetal04}) and the Bohlin (or CALSPEC) sample \citep{Bohletal01,BohlGill04b}, which include a large variety of spectral 
types (from O to M), gravities, and metallicities observed with high accuracy from 1700 \AA\ to 10\,200 \AA. The Vega-based zero points 
in those papers used the \citet{BohlGill04a} SED. Here I present an updated summary which includes the corrections required to adapt the 
results to the \citet{Bohl07} SED.

	For Tycho-2, observed magnitudes for both $V_T$ and $B_T$ are given in the literature, 
thus allowing a direct test of the validity of the sensitivity curves by comparing the observed magnitudes with the synthetic ones derived 
from the STIS spectrophotometry. Also, the Bohlin sample can be used for a direct calculation of the magnitude zero points, yielding
ZP$_{V_T} = 0.036 \pm 0.008$ and ZP$_{B_T} = 0.075 \pm 0.009$. 
However, as reported in \citet{Maiz06a}, an independent measurement for ZP$_{V_T}$ can be obtained from ZP$_V = 0.026\pm 0.008$ (the 
\citealt{BohlGill04a} value, which remains unchanged) and ZP$_{V_T-V} = -0.001 \pm 0.005$, obtained from the NGSL+Bohlin samples, 
to yield ZP$_{V_T} = 0.025 \pm 0.010$. Combining the two using inverse variance weighting we arrive at the final 
value ZP$_{V_T} = 0.032 \pm 0.006$. We can also combine that result with the value for ZP$_{B_T-V_T} = 0.033 \pm 0.005$ from the NGSL+Bohlin
samples to obtain an independent measurement 
for ZP$_{B_T} = 0.065 \pm 0.008$. With the two results for $B_T$ we arrive at a final value ZP$_{B_T} = 0.069 \pm 0.006$.

	For Str\"omgren $u$ and Johnson $U$ \citet{Maiz06a}, discovered that the photometry in the literature was incompatible with the
published sensitivity curves, thus prompting the derivation of new ones. Given the existing confusion between ``old-style'' (energy-integrating)
and ``new-style'' (photon-counting) sensitivity curves, \citet{Maiz06a} also gives the rest of the Str\"omgren and Johnson curves in 
photon-counting form, thus avoiding the possible errors at the hundredth of a magnitude level associated with using the ``old-style'' functions 
in ``new-style'' software. It is important to note that neither of those functions are ``the real sensitivity function'' for those filters,
especially for the case of Johnson $U$. Instead, what they are is an average over the data published in the literature: different observatories
(and observers, atmospheric conditions, and reduction techniques) are likely to yield different effective sensitivity curves, especially to
the left of the Balmer jump, where the atmosphere and not the telescope/filter/detector system is in most cases the main culprit of the
non-detection of photons from the source. 

	For published Str\"omgren and Johnson photometry, typically magnitudes are not given except for $V$. Instead, colors ($b-y$, $B-V$, 
$U-B$) or indices ($m_1$, $c_1$) are provided. Therefore, \citet{Maiz06a} only gives color/index zero points in order to do a consistent 
uncertainty calculation (of course, those can be transformed a posteriori into magnitude zero points). The random uncertainties in the zero 
points are caused by the finite S/N of the data and the natural variation between observing conditions while the systematic errors are 
caused by possible incorrect spectrophotometric or photometric flux calibrations and data reductions. The results, adapted to the new
Vega spectrum, are given in Table~2.

\bigskip

\centerline{{\bf Table 2.} Str\"omgren and Johnson color/index zero points and associated}
\centerline{$\;\;\;\;\;\;\;\;\;\;\;\;$ uncertainties/errors using the new \citet{Bohl07} Vega SED.}

\medskip

\centerline{
\begin{tabular}{lccccccc}
\hline
\hline
            & & \multicolumn{3}{c}{Str\"omgren} & & \multicolumn{2}{c}{Johnson} \\
                \cline{3-5}                         \cline{7-8}
            & & $b-y$ & $m_1$ & $c_1$           & & $B-V$ & $U-B$               \\
\hline
zero point  & & 0.004 & 0.157 & 1.092           & & 0.008 & 0.021               \\
random      & & 0.001 & 0.001 & 0.002           & & 0.001 & 0.006               \\
systematic  & & 0.003 & 0.003 & 0.004           & & 0.004 & 0.014               \\
\hline
\end{tabular}}

\bigskip

\section{Results from NICMOS spectrophotometry}

After \citet{Maiz06a} was published, a number of new spectra were added to the CALSPEC sample \citep{Bohl07}. In particular there are now 23 stars
with NICMOS G096+G141 grism observations which also have 2MASS $JH$ photometry. Of those 23 stars, 14 also have NICMOS G206 observations as well as
2MASS $K_s$ photometry. The NICMOS grism observations have been corrected of the count-rate dependent non-linearity that affects the detector 
(\citealt{Bohl07} and references therein). The stars in the sample span a wide range of temperatures, from hot white dwarfs to red stars.

\bigskip

\centerline{{\bf Table 3.} 2MASS magnitude zero points using the new \citet{Bohl07} Vega SED.}

\medskip

\centerline{
\begin{tabular}{ccc}
\hline
\hline
$J$               & $H$               & $K_s$             \\
\hline
$-0.021\pm 0.005$ & $+0.009\pm 0.005$ & $+0.000\pm 0.006$ \\
\hline
\end{tabular}}

\bigskip

We have applied the technique described in \citet{Maiz05b} to the above sample to measure the zero points for 2MASS $JHK_s$ using the new 
\citet{Bohl07} Vega SED. Results are shown in 
Table 3. The comparison between synthetic and observed magnitudes shows no color terms, indicating that the 2MASS sensitivity curves are 
well characterized. Also, once the zero points have been applied, the distribution of $(m_{{\rm obs},p}-m_{r,p})/\sigma_p$, where $\sigma_p$ is the
photometric uncertainty and $p=J,H,K_s$, is closely approximated by a Gaussian of zero mean and standard deviation of one. This indicates that the 
uncertainties are well behaved and that the largest contribution to the error budget comes from the photometry, not from the spectrophotometry (as it was
the case for the STIS analysis in the prevbious section).

\section{Other recent work on photometric zero points}

Other authors have also recently calculated zero points for commonly used optical/NIR photometric systems. Here I
summarize the work of \citet{HolbBerg06}, \citet{Coheetal03}, and \citet{Cohe07}. They also use a comparison of observed photometry with 
SEDs but with one difference with respect to the work in the previous section: their SEDs include not only observed spectrophotometry but also
atmospheric models. 

	\citet{HolbBerg06} calculate magnitude offsets for Johnson $UBV$, \sindex{Cousins $RI$}, Str\"omgren $uvby$, \sindex{SDSS $ugriz$}, and \sindex{2MASS $JHK_s$}.
Zero points can be immediately calculated by combining them with the observed Vega magnitudes in their Table 1. Their
values are derived from a comparison of observed photometry of white dwarfs with atmospheric models and use the \citet{BohlGill04a}
Vega STIS + 9400 K Kurucz SED. As we show in table 3 and in the next section, their results are in excellent agreement with the filters in common 
with \citet{Maiz06a} with the exception of Str\"omgren $u$ and Johnson $U$. Those differences are easily explainable by their use of the older 
sensitivity curves for those filters, which \citet{Maiz06a} showed to be incorrect (at least in an average sense). 

	\citet{Coheetal03} present the calibration of the 2MASS survey and in that work they give the zero points for $JHK_s$ (which they 
call zero point offsets or ZPOs). It is important to note that their ZPOs are defined with the opposite sign as in Eqn.~\ref{photon} 
and that 
they are evaluated with respect to the \cite{Coheetal92} Vega SED. In Table 4 we present their results adjusted for those effects. Finally,
\citet{Cohe07} presents in these proceedings new ZPs for Tycho-2 $B_TV_T$ derived from their network of FGK dwarfs. Those values are in
good agreement with the ones in the previous section and are also shown in Table 4.

\section{Uniform zero points}

In order to generate a uniform list of magnitude zero points, I have collected the results for the six photometric systems 
(Johnson $UBV$, Cousins $RI$, Str\"omgren $uvby$, Tycho-2 $B_TV_T$,
SDSS $ugriz$, and 2MASS $JHK_s$) mentioned in the previous two sections, transformed color/index zero points into magnitude ones where required,
and adapted them to the new Vega spectrum of \citet{Bohl07} where needed. All of those systems use Vega as the reference SED with the only 
exception of SDSS. The preferred zero points are given in the sixth column of Table 4, with alternate values in the eighth column. The third, 
seventh and ninth columns give the references for the sensitivity curve, preferred zero point, and alternate zero point, respectively, for each 
filter. The fourth and fifth columns give $m_{{\rm ST},p}$(Vega) and $m_{{\rm AB},p}$(Vega), respectively.

	Here are some notes on the results in Table 4. 
The Johnson and Cousins values are ultimately linked to the \citet{BohlGill04a} result of
ZP$_V = 0.026 \pm 0.008$. The possible alternate values for Johnson $U$ and Str\"omgren $u$ from \citet{HolbBerg06} are not given because they 
use a different sensitivity curve, so a direct comparison is not possible.
The Cousins $RI$ sensitivity curves of \citet{Bess83} are in energy-integrating form and should be converted into
photon-counting form before they are used with e.g. {\tt synphot} \citep{synphot}. The preferred Str\"omgren zero points are a combination
of the ZP$_y$ value of \citet{HolbBerg06} with the color/index results of \citet{Maiz06a}. The alternate SDSS zero points are taken from the
SDSS web page and they are of lower precision (one hundredth of a magnitude) than the rest.

\bigskip

\centerline{{\bf Table 4.} ST and AB magnitudes for Vega and magnitude zero points}
\centerline{used by CHORIZOS as of September 2006.}

\medskip

\centerline{
\begin{tabular}{lcccclclc}
\hline
\hline
System           & Filt. & Ref. & $m_{{\rm ST},p}$ & $m_{{\rm AB},p}$ & $\;$ZP$_{r,p}$ & Ref. & $\;$ZP$_{r,p}$ & Ref. \\
                 &       &      & (Vega)           & (Vega)           &                &      & $\;\;\;$alt.   &      \\
\hline
Johnson+Vega     & $U$   & 1    & $-0.160$         & $+0.758$         & $+0.055$       & 1,6  &                &      \\
                 & $B$   & 1    & $-0.603$         & $-0.115$         & $+0.034$       & 1,6  & $+0.027$       & 7    \\
                 & $V$   & 1    & $+0.005$         & $+0.003$         & $+0.026$       & 6    &                & 7    \\
Cousins+Vega     & $R$   & 2    & $+0.547$         & $+0.184$         & $+0.030$       & 7    &                &      \\
                 & $I$   & 2    & $+1.208$         & $+0.419$         & $+0.017$       & 7    &                &      \\
Str\"omgren+Vega & $u$   & 1    & $+0.104$         & $+1.136$         & $+1.432$       & 1,7  &                &      \\
                 & $v$   & 1    & $-0.765$         & $-0.140$         & $+0.179$       & 1,7  & $+0.175$       & 7    \\
                 & $b$   & 1    & $-0.501$         & $-0.156$         & $+0.018$       & 1,7  & $+0.012$       & 7    \\
                 & $y$   & 1    & $+0.006$         & $+0.006$         & $+0.014$       & 7    &                &      \\
Tycho-2+Vega     & $B_T$ & 3    & $-0.650$         & $-0.072$         & $+0.069$       & 1,8  & $+0.079$       & 10   \\
                 & $V_T$ & 3    & $-0.084$         & $-0.017$         & $+0.032$       & 1    & $+0.035$       & 10   \\
SDSS+AB          & $u$   & 4    & $-0.014$         & $+0.922$         & $+0.042$       & 7    & $+0.04$        & 11   \\
                 & $g$   & 4    & $-0.436$         & $-0.106$         & $-0.002$       & 7    & $+0.00$        & 11   \\
                 & $r$   & 4    & $+0.405$         & $+0.143$         & $-0.003$       & 7    & $+0.00$        & 11   \\
                 & $i$   & 4    & $+1.037$         & $+0.356$         & $-0.016$       & 7    & $+0.00$        & 11   \\
                 & $z$   & 4    & $+1.585$         & $+0.518$         & $-0.028$       & 7    & $-0.02$        & 11   \\
2MASS+Vega       & $J$   & 5    & $+2.669$         & $+0.894$         & $-0.021$       & 8    & $-0.014$       & 7    \\
                 &       &      &                  &                  &                &      & $-0.006$       & 9    \\
                 & $H$   & 5    & $+3.763$         & $+1.368$         & $+0.009$       & 8    & $+0.004$       & 7    \\
                 &       &      &                  &                  &                &      & $+0.007$       & 9    \\
                 & $K_s$ & 5    & $+4.823$         & $+1.838$         & $+0.000$       & 8    & $+0.005$       & 7    \\
                 &       &      &                  &                  &                &      & $-0.022$       & 9    \\
\hline
\multicolumn{9}{p{0.90\linewidth}}{\footnotesize  References: [1] \citet{Maiz06a}; [2] \citet{Bess83}; 
[3] \citet{Bess00}; 
[4] {\tt http://www.sdss.org/dr3/instruments/imager/\#filters} as of August 2006;
[5] {\tt http://www.ipac.caltech.edu/2mass/releases/allsky/doc/sec6\_4a.html\#rsr} as of August 2006;
[6] \citet{BohlGill04a}; [7] \citet{HolbBerg06}; [8] this work; [9] \citet{Coheetal03}; 
[10] \citet{Cohe07};
[11] {\tt http://www.sdss.org/dr4/algorithms/fluxcal.html} as of August 2006.}
\end{tabular}}

\bigskip

	Uncertainties for the zero points are not listed in Table 4 because of the heterogeneous character of the sources: some give them, 
some do not, and in some cases the transformation from color to magnitude uncertainties is not obvious given the dependencies between values. 
In most cases where uncertainties are known, they are $\le 1\%$. That precision is consistent with the accuracy derived from a 
one-by-one comparison between the preferred and alternate zero points in Table~4: all but three differ by 0.010 magnitudes or less and only one by
more than 0.016 magnitudes. That exception is 2MASS $K_s$ (0.027 magnitudes between the highest and lowest values), which is notorously difficult 
to calibrate using atmosphere models due to the possible existence of very late M- or early L-type companions \citep{HolbBerg06}. The value 
derived in this paper is located in between the two extremes and is based on observed spectrophotometry, so it should not be affected by such companions,
but is derived from a small sample (14 objects). For that reason, it would be useful to expand the sample of stars observed with NICMOS in order
to confirm the result.

In view of these results, we can conclude that the zero points in Table 4 allow for a comparison between observed photometry and SED models at a level of
accuracy of 1\% or better. 

\section{Implementation in CHORIZOS}

My original motivation when I started working on the issue of zero points was to improve the accuracy of \sindex{CHORIZOS}, an IDL code I 
have worked on for the last five years \citep{Maiz04c}. The purpose of CHORIZOS is to solve the inverse photometric problem using multi-filter 
information with high-precision data. Unfortunately, as the attendees of this conference know well, astronomical photometry has been often 
plagued with data whose accuracy does not match its precision. In other words, it makes little sense to use photometric uncertainties 
measured in milimagnitudes when the systematic errors introduced by erroneous zero points are measured in hundredths of magnitudes. Hopefully,
time will demonstrate that the zero points presented here do indeed provide the desired 1\% accuracy. They were included in CHORIZOS starting
with version 2.1.3 in September 2006. Previous versions of the software had more inaccurate values (up to 2\%), so pasts users may want to
consider reprocessing their data. We are currently in the process of checking the zero points described in this paper using a sample of low-extinction
OB stars with multi-filter photometry analyzed with CHORIZOS; our preliminary results indicate that they are indeed accurate to within 1\%.

	In its current (October 2006) version, CHORIZOS calculates synthetic colors from a grid of model spectral energy distributions and 
compares them with an observed set of magnitudes for an object to select which models are compatible with the 
observations. Several stellar and cluster SEDs and 88 filters are preinstalled but the user can introduce his/her own. The user may also select
the type and amount of extinction as two additional parameters and change the range and spacing of the SED grid. The multifilter approach has
several important advantages with respect to the traditional color-magnitude or color-color ones. First, processing all the available 
information simultaneously allows for the easy detection of erroneous data (e.g. an incorrect magnitude) or of objects than do not conform to the
models (e.g. binary stars in a single-star sample or objects outside the assumed parameter range). Also, the combination of 
multi-filter information with a large wavelength coverage makes the simultaneous fit of several parameters (temperature, extinction, age, 
gravity\ldots) possible, thus permitting the reduction on the number of a priori assumptions on the properties of the object.
Finally, since CHORIZOS provides most of its output (text and graphical) in parameter space (as opposed to color or magnitude space), it 
allows for the easy calculation of the properties of the object studied, including the possible existence of multiple solutions compatible with
the observed data.

	My plans for the future two years include the following modifications to the code:

\begin{itemize}
  \item A change of the comparison mechanism between synthetic and observed data from colors to magnitudes. This will not impact the previously
	calculated results but will allow several other changes, such as adding distance or redshift as an extrinsic parameter.
  \item Inclusion of spectrophotometric data as input: absolute fluxes, ratios, indices, and equivalent widths.
  \item Inclusion of multicolor indices (Str\"omgren-like) as input for photometric data.
  \item Possibility of having detection limits instead of measured magnitudes.
  \item Use of arbitrary wavelength gids.
  \item Addition of full Bayesian priors.
  \item Addition of more graphical capabilities.
  \item Inclusion of an algorithm to search and model multiple solutions.
  \item Design of a web-based interface and integration with STSDAS {\tt synphot}.
\end{itemize}

	In addition, it is possible that small ($\le 1\%$) revisions to the zero points will be made if new data becomes available. CHORIZOS can 
be downloaded from {\tt http://www.stsci.edu/\~{}jmaiz}.

\acknowledgements  

I would like to thank Chris Sterken for organizing such an opportune meeting and for his understanding during the edition process. 
I also greatly appreciated the useful information 
exchanges before, during, and after the conference with Martin Cohen, Jay Holberg, and Ralph Bohlin (among others), which made
this contribution possible.

\bibliographystyle{astron}

\begin{thebibliography}{}

\bibitem[\protect\astroncite{{Aufdenberg} et~al.}{2006}]{Aufdetal06}{Aufdenberg}, J.~P., et~al. 2006,
\newblock {\apj \ } {645}, 664

\bibitem[\protect\astroncite{{Bessell}}{1983}]{Bess83}{Bessell}, M.~S. 1983,
\newblock {\pasp \ } {95}, 480

\bibitem[\protect\astroncite{{Bessell}}{2000}]{Bess00}{Bessell}, M.~S. 2000,
\newblock {\pasp \ } {112}, 961

\bibitem[\protect\astroncite{{Bohlin}}{2007}]{Bohl07}{Bohlin}, R.~C. 2007,
\newblock {\em these Proceedings}

\bibitem[\protect\astroncite{{Bohlin} et~al.}{2001}]{Bohletal01}{Bohlin}, R.~C., {Dickinson}, M.~E., \& {Calzetti}, D. 2001,
\newblock {\aj \ } {122}, 2118

\bibitem[\protect\astroncite{{Bohlin} \& {Gilliland}}{2004a}]{BohlGill04b}{Bohlin}, R.~C. \& {Gilliland}, R.~L. 2004a,
\newblock {\aj \ } {128}, 3053

\bibitem[\protect\astroncite{{Bohlin} \& {Gilliland}}{2004b}]{BohlGill04a}{Bohlin}, R.~C. \& {Gilliland}, R.~L. 2004b,
\newblock {\aj \ } {127}, 3508

\bibitem[\protect\astroncite{{Cohen}}{2007}]{Cohe07}{Cohen}, M. 2007,
\newblock {\em these proceedings}

\bibitem[\protect\astroncite{{Cohen} et~al.}{1992}]{Coheetal92}{Cohen}, M., {Walker}, R.~G., {Barlow}, M.~J., \& {Deacon}, J.~R. 1992,
\newblock {\aj \ } {104}, 1650

\bibitem[\protect\astroncite{Cohen et~al.}{2003}]{Coheetal03}Cohen, M., Wheaton, W.~A., \& Megeath, S.~T. 2003,
\newblock {\aj \ } {126}, 1090

\bibitem[\protect\astroncite{Goudfrooij et~al.}{2006}]{Goudetal06}Goudfrooij, P., Bohlin, R.~C., Ma\'{\i}z~Apell\'aniz, J., \& Kimble, R.~A.
  2006,
\newblock {\pasp \ } October issue

\bibitem[\protect\astroncite{{Gregg} et~al.}{2004}]{Gregetal04}{Gregg}, M.~D., et~al. 2004,
\newblock {\em American Astronomical Society Meeting Abstracts} 205

\bibitem[\protect\astroncite{{Hayes}}{1985}]{Haye85}{Hayes}, D.~S. 1985,
\newblock in {IAU Symp. 111: Calibration of Fundamental Stellar
  Quantities}, 225

\bibitem[\protect\astroncite{{Holberg} \& {Bergeron}}{2006}]{HolbBerg06}{Holberg}, J.~B. \& {Bergeron}, P. 2006,
\newblock {\aj \ } {132}, 1221

\bibitem[\protect\astroncite{Laidler et~al.}{2005}]{synphot}Laidler, V. et~al. 2005,
\newblock { {Synphot User's Guide v5.0 (STScI: Baltimore)}}

\bibitem[\protect\astroncite{Ma\'{\i}z~Apell\'aniz}{2004}]{Maiz04c}Ma\'{\i}z~Apell\'aniz, J. 2004,
\newblock {\pasp \ } {116}, 859

\bibitem[\protect\astroncite{Ma\'{\i}z~Apell\'aniz}{2005}]{Maiz05b}Ma\'{\i}z~Apell\'aniz, J. 2005,
\newblock {\pasp \ } {117}, 615

\bibitem[\protect\astroncite{Ma\'{\i}z~Apell\'aniz}{2006}]{Maiz06a}Ma\'{\i}z~Apell\'aniz, J. 2006,
\newblock {\aj \ } {131}, 1184

\bibitem[\protect\astroncite{{M\'egessier}}{1995}]{Mege95}{M\'egessier}, C. 1995,
\newblock {\aap \ } {296}, 771

\bibitem[\protect\astroncite{{Peterson} et~al.}{2006}]{Peteetal06}{Peterson}, D.~M., et~al. 2006,
\newblock {\nat \ } {440}, 896

\end{thebibliography}

\question{Ivezic} All your zeropoint corrections have the same sign. Can you exclude the possibility that the STIS calibration is off by $\sim 2$\% in the relevant $\lambda$ range?

\answer{Ma\'{\i}z Apell\'aniz} It is true that the zeropoint corrections for $b-y, B-V, U-B$, and $B_T-V_T$ are all positive (0.004, 0.008, 0.021, and 0.037 magnitudes, respectively). However, I do not think
this is a systematic STIS effect for two reasons:  
First, the uncertainty introduced by the WD calibration (see talk by Ralph Bohlin) is expected to be only 0.002--0.004 magnitudes at most.
Second, different authors (M. Cohen, J. Holberg) have recently measured consistent results using non-STIS data.

\question{Sterken} Related to Bessel's remark: the deviations in $c_1$  were described by us in the LTPV-project as conformity errors. We have used up to 13 different $uvby$ systems and we found,
empirically, that several such systems were not transformable. It would be useful to run your analysis on our results.

\answer{Ma\'{\i}z Apell\'aniz} I agree that is one possible explanation for the effects I see. Indeed, it would be a good idea to test those 13 different systems to check it out.


\end{document}